\documentclass[twoside,fleqn,a4paper]{article}
\usepackage{espcrc2MOD}
\usepackage{epsfig}

\newcommand{\ga}{\gamma}

\newcommand{\varep}{\varepsilon}

\newcommand{\la}{\lambda}
\newcommand{\rh}{\rho}

\newcommand{\si}{\sigma}

\newcommand{\varph}{\varphi}

\newcommand{\ps}{\psi}

\newcommand{\vac}[1]{\mbox{\boldmath$\langle$}{#1}\mbox{\boldmath$\rangle$}}
\newcommand{\klgl}{\:\hbox to -0.2pt{\lower2.5pt\hbox{$\sim$}\hss}
 {\raise3pt\hbox{$<$}}\:}
\newcommand{\fm}{{\rm \;fm}}
\newcommand{\GeV}{{\rm \;GeV}}
\newcommand{\nn}{\nonumber}
\newcommand{\figeps}[2]{\epsfxsize=#1mm\epsfbox[72 240 540 540]{./#2.eps}}
\newcommand{\figepsSC}[2]{\epsfxsize=#1mm\epsfbox[72 240 540 540]{./#2.eps}
  \setlength{\unitlength}{0.464mm}}
\newcommand{\figpsSC}[2]{\epsfxsize=#1mm\epsfbox[100 300 530 570]{./#2.ps}
  \setlength{\unitlength}{0.464mm}}
\begin{document}
\bibliographystyle{personale}
\sloppy
\begin{titlepage}
\title{{\Large\bf Exclusive production of vector mesons
  \mbox{\boldmath $\rho$, $\rho'$} and \mbox{\boldmath$\rho''$} by real and virtual photons}
  \thanks{Talk presented in Montpellier, QCD Euroconference 98}}
\author{G.~Kulzinger \thanks{Supported by the Deutsche Forschungsgemeinschaft
  under grant no. GRK 216/1-96}$^{,}$
\address{Institut f\"ur Theoretische Physik der Universit\"at Heidelberg,\\
  Philosophenweg 16, D-69120 Heidelberg\\
  E-mail: G.Kulzinger@thphys.uni-heidelberg.de}}
\begin{abstract}
  We present a non-perturbative QCD calculation of high-energy diffractive photo- and leptoproduction of vector mesons $\rh$, $\rh'$ and $\rh''$ on a nucleon. The initial photon splits up in a $q\bar{q}$-dipole and transforms into a vector meson by scattering on the quark-diquark nucleon. The dipole-dipole scattering amplitude is provided by the non-perturbative model of the stochastic QCD vacuum, the wave functions result from considerations on the light-cone. We assume the physical $\rh'$- and $\rh''$-states to be mixed states of an active $2S$-excitation and a rest whose coupling to the photon is suppressed. We obtain good agreement with the experimental data and get an understanding of the markedly different spectrum in the $\pi^+\pi^-$-invariant mass for photoproduction and $e^+e^-$-an\-nihilation.
\end{abstract}
\end{titlepage}
\maketitle
\vspace*{-10cm}
\noindent\begin{minipage}{16cm}\begin{center}
  HD-THEP-98-37. -- {\it To be published in} Nucl.Phys.{\bf B} (Proc.Suppl.)
\end{center}\end{minipage}
\vspace*{8.3cm}
\section{Introduction}
Diffractive scattering processes are characterized by small momentum transfer, $t\klgl 1$~GeV$^2$, and thus governed by non-perturbative QCD. To get more insight in the physics at work we inves\-tigate exclusive vector meson production by real and virtual photons. In this note we summarize recent results from Ref.~\cite{KDP} on $\rh$-, $\rh'$- and $\rh''$-pro\-duction. In Refs~\cite{KDP,DGKP} we have developed a framework which we here can only flash.

We consider high-energy diffractive collision of a photon which dissociates into a $q\bar{q}$-dipole and transforms into a vector meson with a proton in the quark-diquark picture which remains intact. The scattering $T$-amplitude can be written as an integral of the dipole-dipole amplitude and the corresponding wave functions. Integrating out the proton side, we have
\begin{eqnarray} \label{T_amplitude}
T^\la_V(s,t) = is \int \frac{dzd^2{\bf r}}{4\pi}\,
  \psi^\dagger_{V(\la)}\psi_{\ga(Q^2,\la)}(z,{\bf r}) \hspace*{1em}&& \nn \\
\times J_p(z,{\bf r},\Delta_T) \;, &&
\end{eqnarray}
where $V(\la)$ stands for the final vector meson and $\ga(Q^2,\la)$ for the initial photon with definite helicities $\la$ (and virtuality $Q^2$); $z$ is the light-cone momentum fraction of the quark, $\bf r$ the transverse extension of the $q\bar{q}$-dipole. The function $J_p(z,{\bf r},\Delta_T)$ is the interaction amplitude for a dipole $\{z,{\bf r}\}$ scattering on a proton with fixed momentum transfer $t\!=\!-\Delta_T^2$; for $\Delta_T\!=\!0$ due to the optical theorem it is the corresponding total coss section (see below Eq.~(\ref{sigma_dipole})). It is calculated within non-perturbative QCD: In the high-energy limit Nachtmann~\cite{Na} derived a non-perturbative formula for dipole-dipole scattering whose basic entity is the vacuum expectation value of two lightlike Wilson loops. This gets evaluated in the model of the stochastic QCD vacuum~\cite{DFK}.

\section{The model of the stochastic vacuum}
Coming from the functional integral approach the model of the stochastic QCD vacuum~\cite{DoSi} assumes that the non-perturbative part of the gauge field measure, i.e. long-range gluon fluctuations that are associated with a non-trivial vacuum structure of QCD, can be approximated by a stochastic process in the gluon field strengths with convergent cumulant expansion. Further assuming this process to be gaussian one arrives at a description through the second cumulant $\vac{g^2 F_{\mu\nu}^A(x;x_0) F_{\rh\si}^{A'} (x';x_0)}$ which has two Lorentz tensor structures multiplied by correlation functions $D$ and $D_1$, respectively. $D$ is non-zero only in the non-abelian theory or in the abelian theory with magnetic monopoles and yields linear confinement. Whereas the $D_1$-structure is not confining.

\begin{figure}
$$
\figpsSC{65}{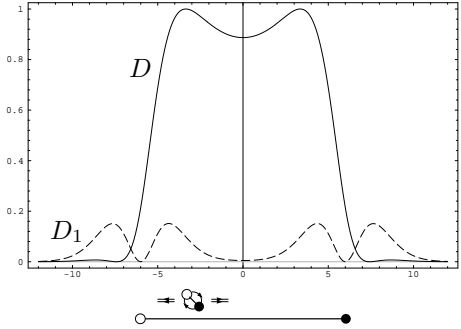}
\begin{picture}(0,0)
\put(-124,24){\makebox(0,0){$D_1$}}
\put(-103,71){\makebox(0,0){$D$}}
\end{picture}
$$
\vspace*{-9ex}
\caption[]{\footnotesize Interaction amplitude (arbitrary units) of two colour dipoles as function of their impact (units of correlation lengths $a$). One large $q\bar{q}$-dipole of extension $12a$ is fixed, the second small one of extension $1a$ is, averaged over all its orientations, shifted along on top of the first one. For the $D_1$-tensor structure of the correlator there are only contributions when the endpoints are close to each other, whereas for the $D$-structure large contributions show up also from between the endpoints. This is to be interpreted as interaction with the gluonic string between the quark and antiquark.} \label{Fig:dipdip}
\end{figure}

The underlying mechanism of (interacting) gluonic strings also shows up in the scattering of two colour dipoles, cf.~Fig.~\ref{Fig:dipdip}, and essentially determines the $T$-amplitude if large dipole sizes are not suppressed by the wave functions. To confront with experiment this specific-large distance prediction we are intended to study the broad $\rh$-states and, especially, their production by broad small-$Q^2$ photons. Before we enter the discussion of our results, however, we have to specify these states and have to fix their wave functions as well as that of the photon.
%
%
\vspace*{-1.5mm}
\section{Physical states \mbox{\boldmath $\rho$, $\rho'$} and \mbox{\boldmath$\rho''$}}
\begin{figure}[!b]
$$
\figpsSC{65}{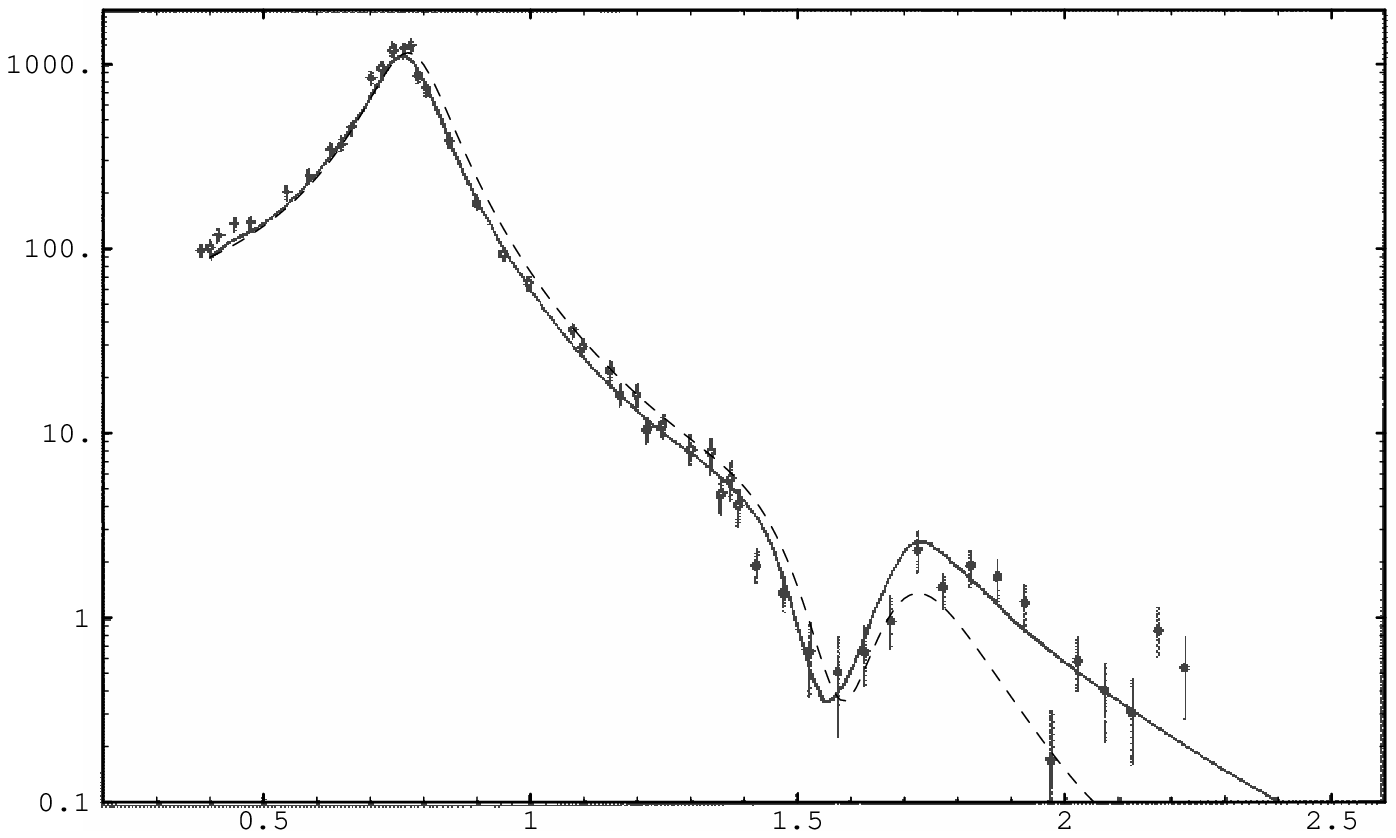}
\begin{picture}(0,0)
\put(-32,78){\makebox(0,0){\scriptsize signs pattern}}
\put(-32,72){\makebox(0,0){\scriptsize \{+,--,+\}}}
\put(-2,3){\makebox(0,0){$M[\!\!\GeV]$}}
\put(-140,89){\makebox(140,0){$\si(M)[nb]$
    \hfill $e^+e^-\rightarrow\pi^+\pi^-$}}
\end{picture}
$$
\vspace*{-3ex}
$$
\figpsSC{65}{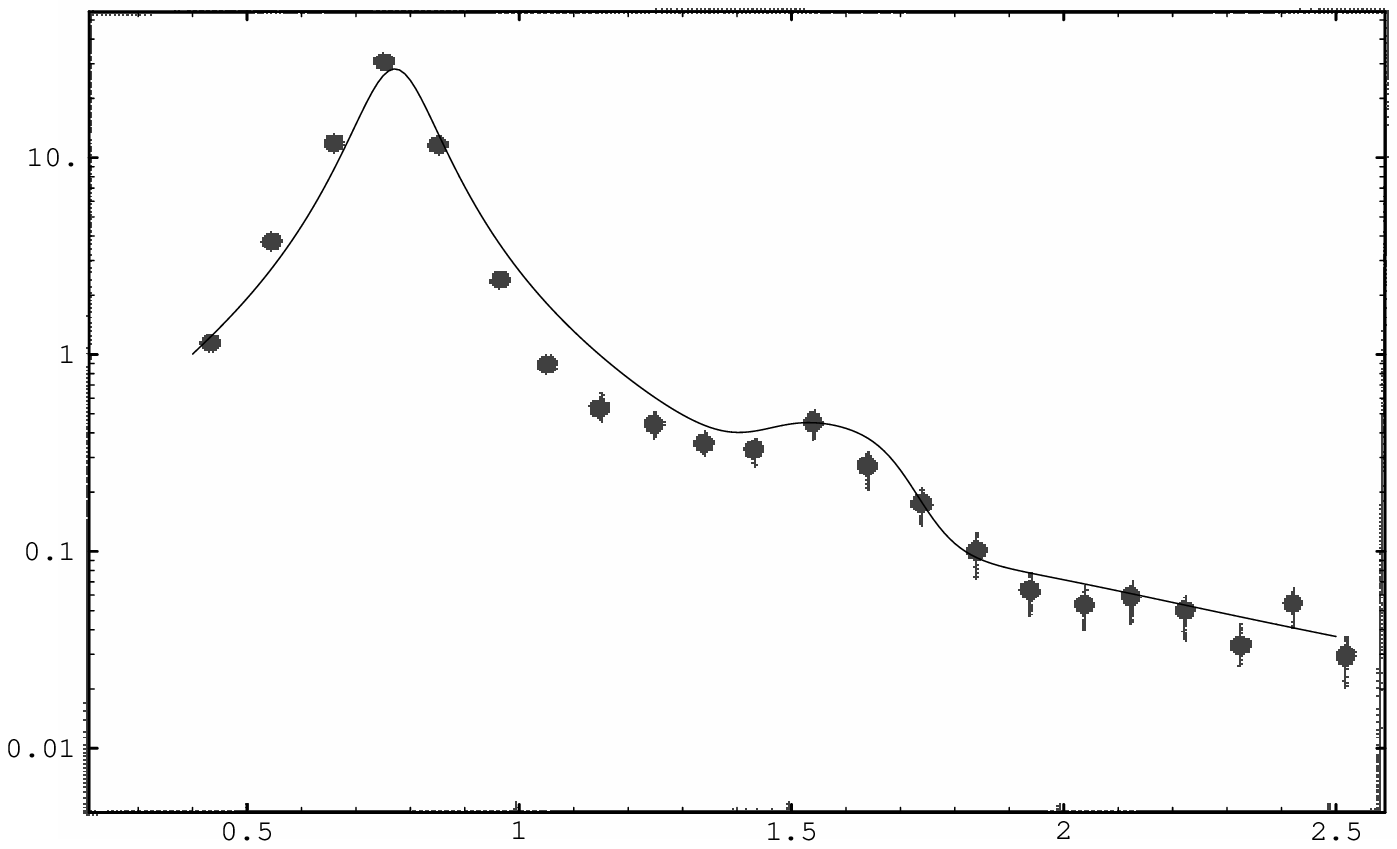}
\begin{picture}(0,0)
\put(-32,79){\makebox(0,0){\scriptsize signs pattern}}
\put(-32,73){\makebox(0,0){\scriptsize \{+,+,--\}}}
\put(-2,3){\makebox(0,0){$M[\!\!\GeV]$}}
\put(-140,90){\makebox(135,0){$d\si\!/\!dM(M)[\mu b\!/\!\!\GeV]$
    \hfill $\ga p\rightarrow\pi^+\pi^-p$}}
\end{picture}
$$
\vspace*{-11ex}
\caption[]{\footnotesize Mass spectra of $e^+e^-$-annihilation into $\pi^+\pi^-$ and $\pi^+\pi^-$-photoproduction on the proton: The interference in the 1.6 GeV region is destructive and constructive, respectively, and shows specific amplitude signs patterns.} \label{Fig:spectra}
\end{figure}

Analyzing the $\pi^+\pi^-$-invariant mass spectra for photoproduction and $e^+e^-$-annihilation Donnachie and Mirzaie~\cite{DoMi} concluded evidence for two resonances in the 1.6~GeV region whose masses are compatible with the $1^{--}$ states $\rh(1450)$ and $\rh(1700)$, cf. Fig.~\ref{Fig:spectra}. We make as simplest ansatz
\begin{eqnarray}
|\rho(770)\rangle\;\, &=& \;\;\; |1S\rangle \;, \nn \\
|\rho(1450)\rangle    &=& \;\;\; \cos\theta \; |2S \rangle
                               + \sin\theta \; |rest \rangle \;, \nn \\
|\rho(1700)\rangle    &=&       -\sin\theta \; |2S \rangle
                               + \cos\theta \; |rest \rangle \;,
\end{eqnarray}
where $|rest\rangle$ is considered to have $|2D\rangle$- and hybrid component whose couplings to the photon both are suppressed, see Ref.~\cite{ClDo,ClPa}. With our convention of the wave functions the relative signs $\{+,-,+\}$ of the production amplitudes of the $\rh$-, $\rh'$- and $\rh''$-states in $e^+e^-$-annihilation determine the mixing angle as $\theta\!=\!41.2^\circ$. With this value and the branching ratios in $\pi^+\pi^-$ extracted in Ref.~\cite{DoMi} we calculate the photoproduction spectrum as shown in Fig.~\ref{Fig:spectra} with the observed signs pattern $\{+,+,-\}$, for details cf.~\cite{KDP}. We will~understand below in Fig.~\ref{Fig:overlap2S} the sign change of the $2S$-production as due to the dominance of large dipole sizes in photoproduction in contrary to the coupling to the electromagnetic current $f_{2S}$ being determined by small dipole sizes.
%
%
\vspace*{-3mm}
\section{Light-cone wave functions}
In the high-energy limit the photon can be identified as its highest Fock $q\bar{q}$-component. The vector meson wave function arises by distributing this $q\bar{q}$-dipole $\{z,{\bf r}\}$.

{\bf Photon.} With mean of light-cone perturba\-tion theory (LCPT) we get explicit expressions for both longitudinal and transverse photons. The photon transverse size which we will see to determine the $T$-amplitude is governed by the product $\varep r$, $\varep \!=\! \sqrt{z\bar{z} Q^2 \!+\! m^2}$. For high $Q^2$ longitudinal photons dominate by a power of $Q^2$; their $z$-endpoints being explicitly suppressed, LCPT is thus applicable. For moderate $Q^2$ also transverse photons contribute which have large extensions because endpoints are not suppressed. For $Q^2$ smaller than 1~GeV$^2$ LCPT definitively breaks down. However, it was shown~\cite{DGP} that a quark mass phenomenologically interpolating between a zero valence and a $220$~MeV constituent mass astonishing well mimics chiral symmetry breaking and confinement. Our wave function is thus given by LCPT with such a quark mass $m(Q^2)$, for details cf. Refs~\cite{KDP,DGKP}.

{\bf Vector mesons.} The vector mesons wave functions of the $1S$- and $2S$-states are modelled according to the photon. We only replace the photon energy denominator $(\varep^2\!+\!k^2)^{-1}$ by a function of $z$ and $k$ for which ans\"atze according to Wirbel and Stech~\cite{WSB} are made; for the "radial" excitation we account by both a polynomial in $z\bar{z}$ and the $2S$-polynomial in $k^2$ of the transverse harmonic oscillator. The parameters are fixed by the demands that the $1S$-state reproduces $M_\rh$ and $f_\rh$ and the $2S$-state is both normalized and orthogonal on the $1S$-state. Details again in Ref.~\cite{KDP}.

\section{Results}
Before presenting our results we remark that all calculated quantities are absolut predictions. The cross sections are constant with total energy $s$ due to the eikonal approximation applied and refer to $\sqrt{s}\!=\!20$~GeV where the proton radius as the third parameter of the model of the stochastic QCD vacuum is fixed; its two other parameters, the gluon condensate $\vac{g^2FF}$ and the correlation length $a$, are determined by matching low-energy and lattice results, cf. Ref.~\cite{DFK}.

%
%
\begin{figure}
$$
\figepsSC{64}{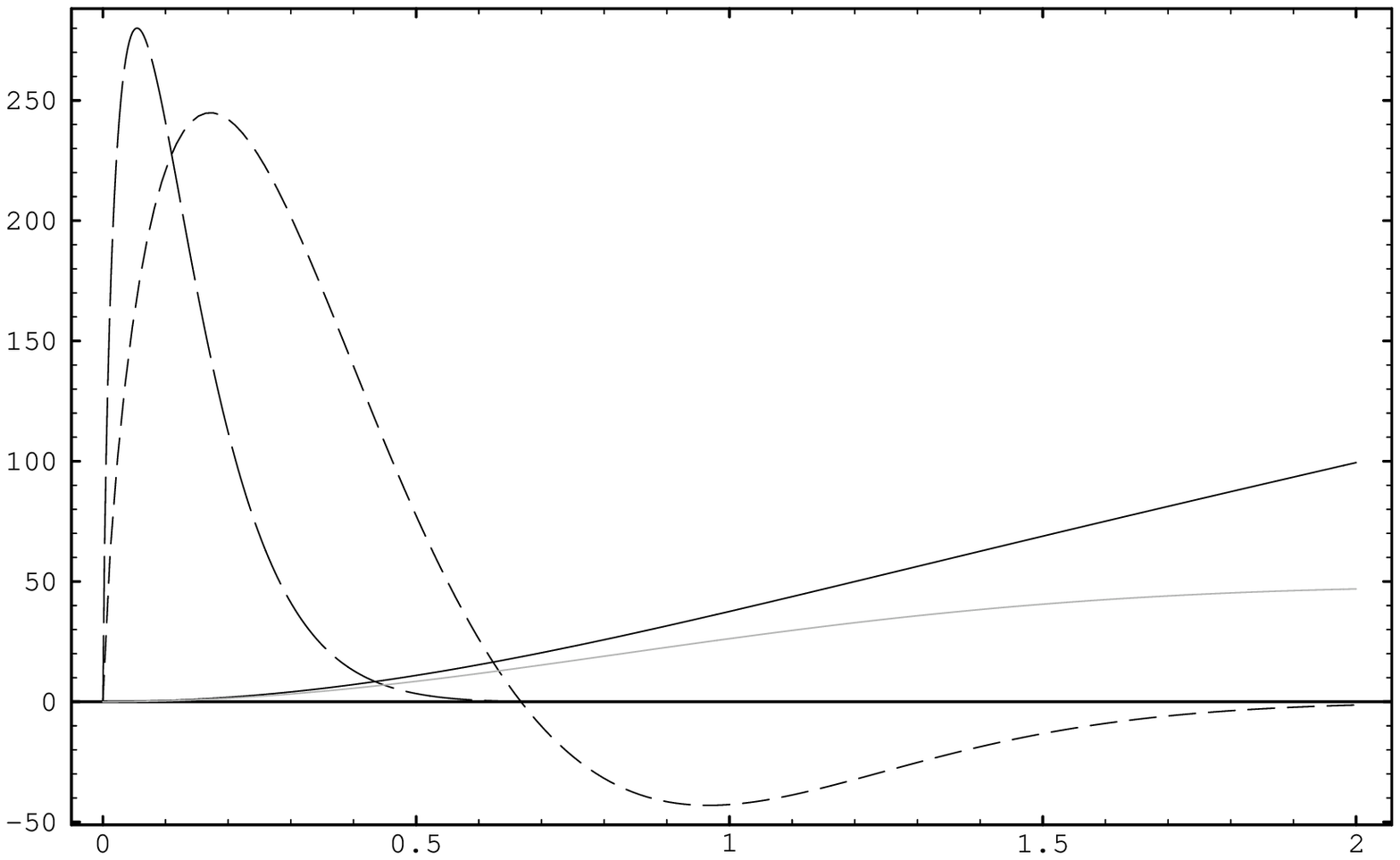}
\begin{picture}(0,0)
\put(2,0){\makebox(0,0){$r[\!\!\fm]$}}
\put(-152,91){\makebox(165,10){
    $r\ps_{V(\la)}^\dagger\ps_{\ga(Q^2,\la)}(r)[10^{-3}\fm^{-1}]$
    \hfill $J_p^{(0)}(1\!/\!2,r)[\rm mb]$}}
\put(-62,82){\makebox(0,0){2S-Longitudinal}}
\put(-87,60){\makebox(0,0){\scriptsize $Q^2\!=\!1\GeV^2$}}
\put(-112,84){\makebox(0,0){\scriptsize $Q^2\!=\!20\GeV^2$}}
\end{picture}
$$
\vspace*{-3ex}
$$
\figepsSC{65}{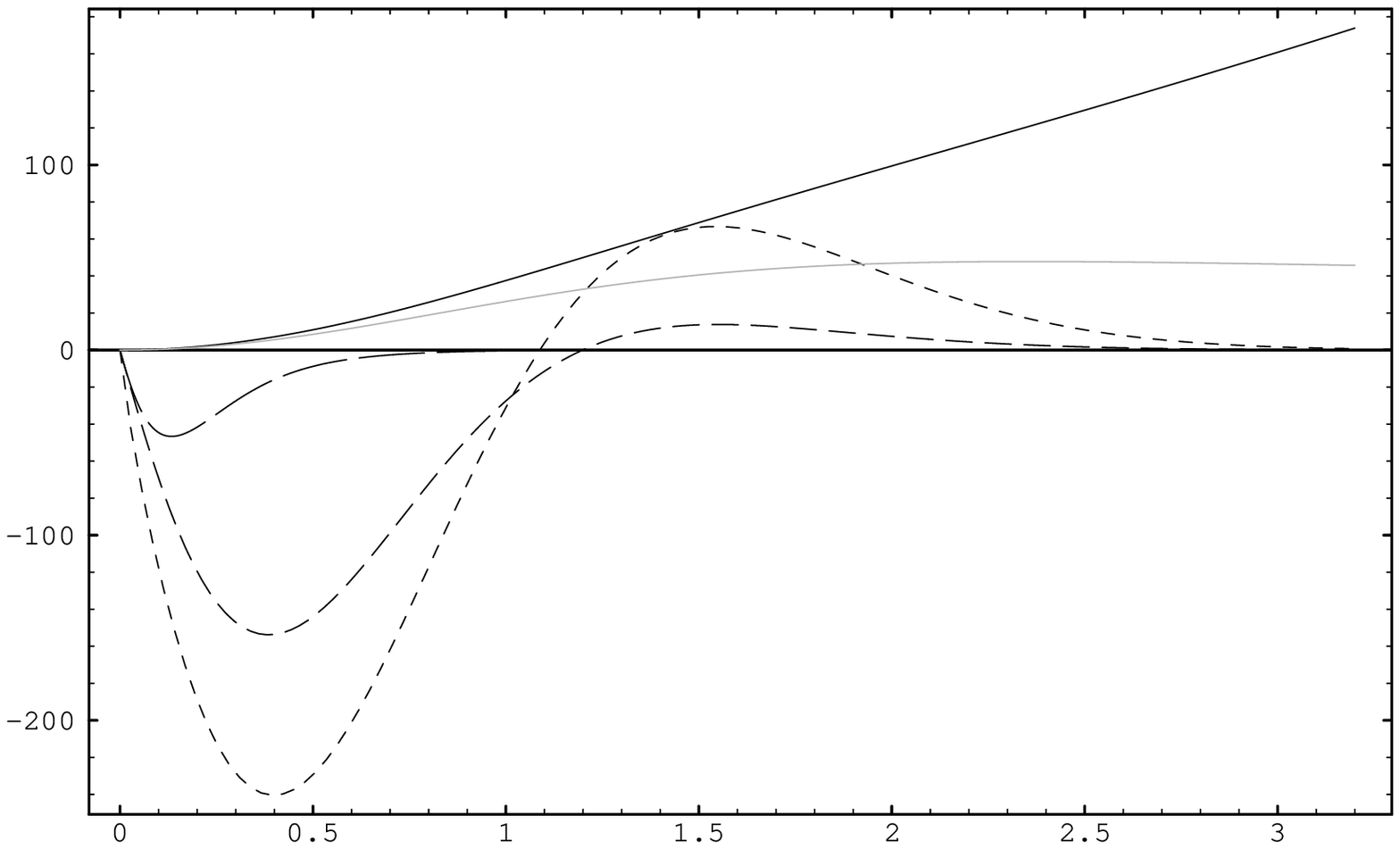}
\begin{picture}(0,0)
\put(2,1){\makebox(0,0){$r[\!\!\fm]$}}
\put(-62,82){\makebox(0,0){2S-Transverse}}
\put(-97,12){\makebox(0,0){\scriptsize $Q^2\!=\!0$}}
\put(-111,38){\makebox(0,0){\scriptsize $Q^2\!=\!1\GeV^2$}}
\put(-100,47){\makebox(0,0){\scriptsize $Q^2\!=\!20\GeV^2$}}
\end{picture}
$$
\vspace*{-9.5ex}
\caption[]{\footnotesize Dipole-proton total cross section $J_p^{(0)}$ and the effective overlap $r\ps_{V(\la)}^\dagger\ps_{\ga(Q^2,\la)}$ as function of the transverse dipole size $r$. The black lines are the function \mbox{$J_p^{(0)}(1\!/\!2, r)$}, Eq.~(\ref{sigma_dipole}), i.e. the total cross section of a $q\bar{q}$-dipole $\{z\!=\!1\!/\!2,{\bf r}\}$, averaged over all orientations, scattering on a proton; the grey lines show the cross section of a completely abelian, non-confining theory. The $T$-amplitude is obtained by integration over the product of $J_p$ and the overlap function, which essentially, cf. Eq.~(\ref{overlap}), is the quantity shown for $Q^2\!=\!0,\;1$ and 20~GeV$^2$ as short, medium and long dashed curves, respectively. } \label{Fig:overlap2S}
\end{figure}
In Fig.~\ref{Fig:overlap2S} we display both the functions
\begin{equation} \label{sigma_dipole}
J_p^{(0)}(z,r) :=
  \int_0^{2\pi} \frac{d\varph_{\bf r}}{2\pi}\,J_p(z,{\bf r},\Delta_T=0)
\end{equation} 
and
\begin{eqnarray} \label{overlap}
\hspace*{-2em}&&r\ps_{V(\la)}^\dagger\ps_{\ga(Q^2,\la)}(r) \nn \\
\hspace*{-2em}&&:= \int \frac{dz}{4\pi}\; \int_0^{2\pi} \frac{d\varph_{\bf r}}{2\pi}\;
       |{\bf r}|\,\ps_{V(\la)}^\dagger\ps_{\ga(Q^2,\la)}(z,{\bf r})
\end{eqnarray}
which together, according to Eq.~(\ref{T_amplitude}), determine the leptoproduction amplitude. The curve for photoproduction of the transverse state strikingly shows how the outer positive region of the wave functions effective overlap $r\ps_{V(\la)}^\dagger\ps_{\ga(Q^2,\la)}(r)$ wins over the inner negative part due to the strong rise with $r$ of the dipole-proton interaction amplitude $J_p^{(0)}$. This rise itself is a consequence of the string interaction mechanism discussed above. Note, that to the cross section dipole sizes up to $2.5$~fm contribute significantly.

\begin{figure}[!t]
\begin{minipage}{160mm}
\hspace*{-0.5mm}\begin{minipage}{80mm}
\setlength{\unitlength}{1mm}
\begin{picture}(80,54.5)(0,0)
\put(0,0){ \makebox(80,49.5){\figeps{76.5}{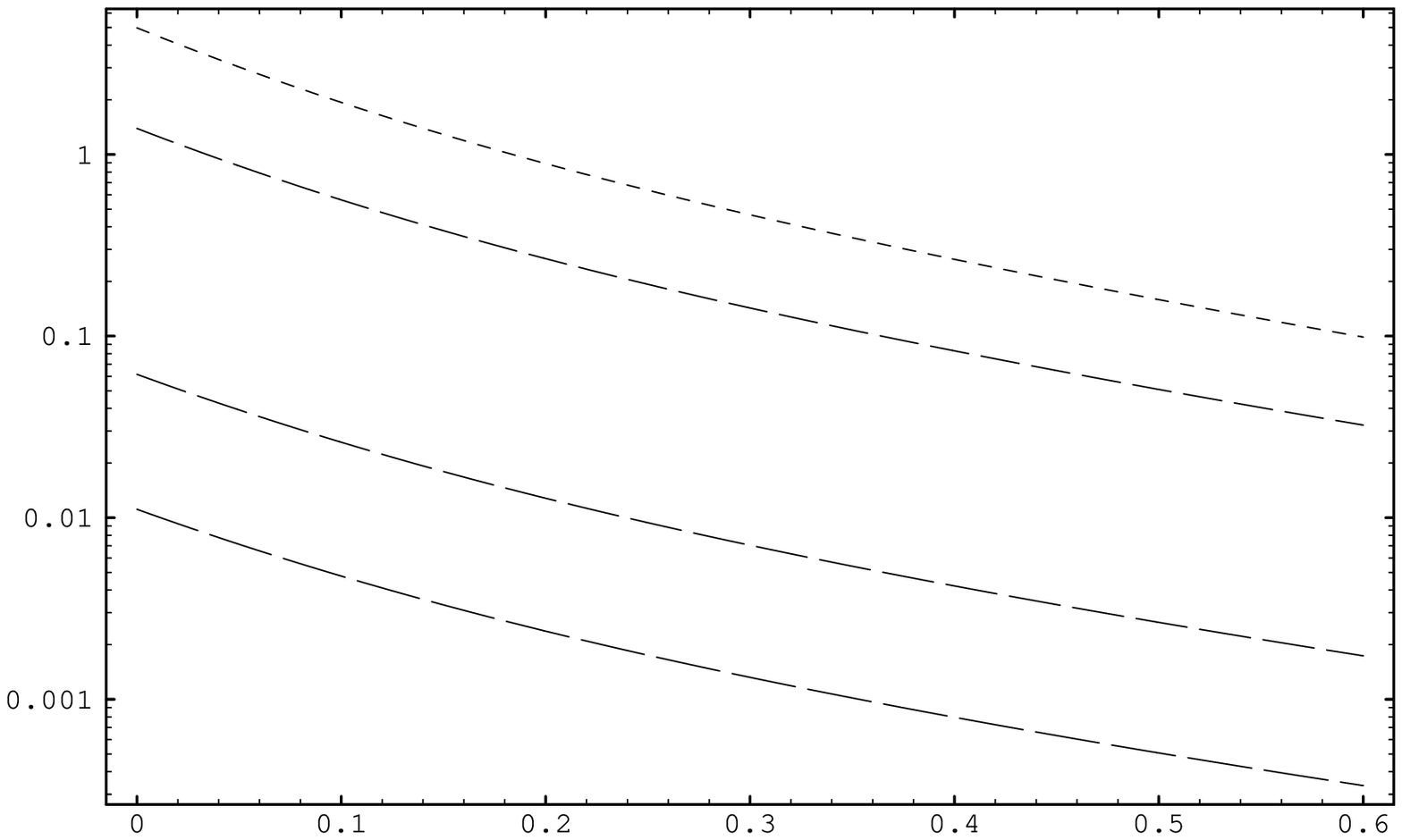}} }
  \put(17,52){\makebox(0,0){
      $d\si\!/\!dt[\mu b \!\times\!\!\!\GeV^{-2}]$}}
  \put(67.75,45.4){\makebox(0,0){$\rh$-Longitudinal}}
  \put(40,42){\makebox(0,0){\scriptsize $Q^2\!=\!1\!/\!4\GeV^2$}}
  \put(19.75,34.5){\makebox(0,0){\scriptsize $Q^2\!=\!2\GeV^2$}}
  \put(40,23.5){\makebox(0,0){\scriptsize $Q^2\!=\!10\GeV^2$}}
  \put(20.25,13.5){\makebox(0,0){\scriptsize $Q^2\!=\!20\GeV^2$}}
\end{picture}
\end{minipage}
\hspace*{-2mm}\begin{minipage}{81mm}
\setlength{\unitlength}{1mm}
\begin{picture}(80,54.5)
\put(0,0){ \makebox(80,49.5){\figeps{78}{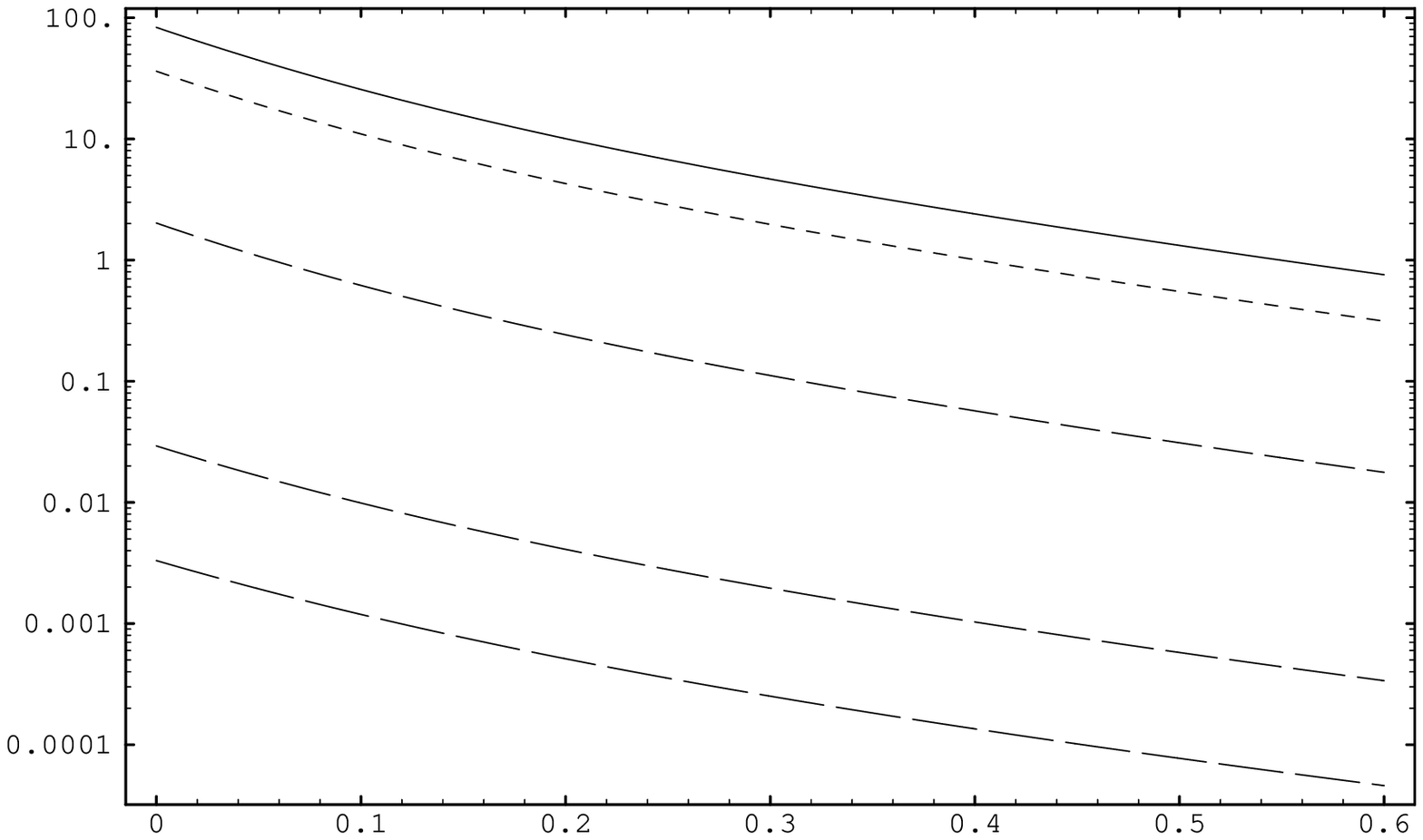}} }
  \put(70.5,45.75){\makebox(0,0){$\rh$-Transverse}}
  \put(38.5,43){\makebox(0,0){\scriptsize $Q^2\!=\!0$}}
  \put(43,34){\makebox(0,0){\scriptsize $Q^2\!=\!1\!/\!4\GeV^2$}}
  \put(19,30.75){\makebox(0,0){\scriptsize $Q^2\!=\!2\GeV^2$}}
  \put(43,20.5){\makebox(0,0){\scriptsize $Q^2\!=\!10\GeV^2$}}
  \put(19.75,12.25){\makebox(0,0){\scriptsize $Q^2\!=\!20\GeV^2$}}
\end{picture}
\end{minipage}
\hspace*{-0.5mm}\begin{minipage}{80mm}
\setlength{\unitlength}{1mm}
\begin{picture}(80,49.5)
\put(-1.25,0){ \makebox(80,49.5){\figeps{80}{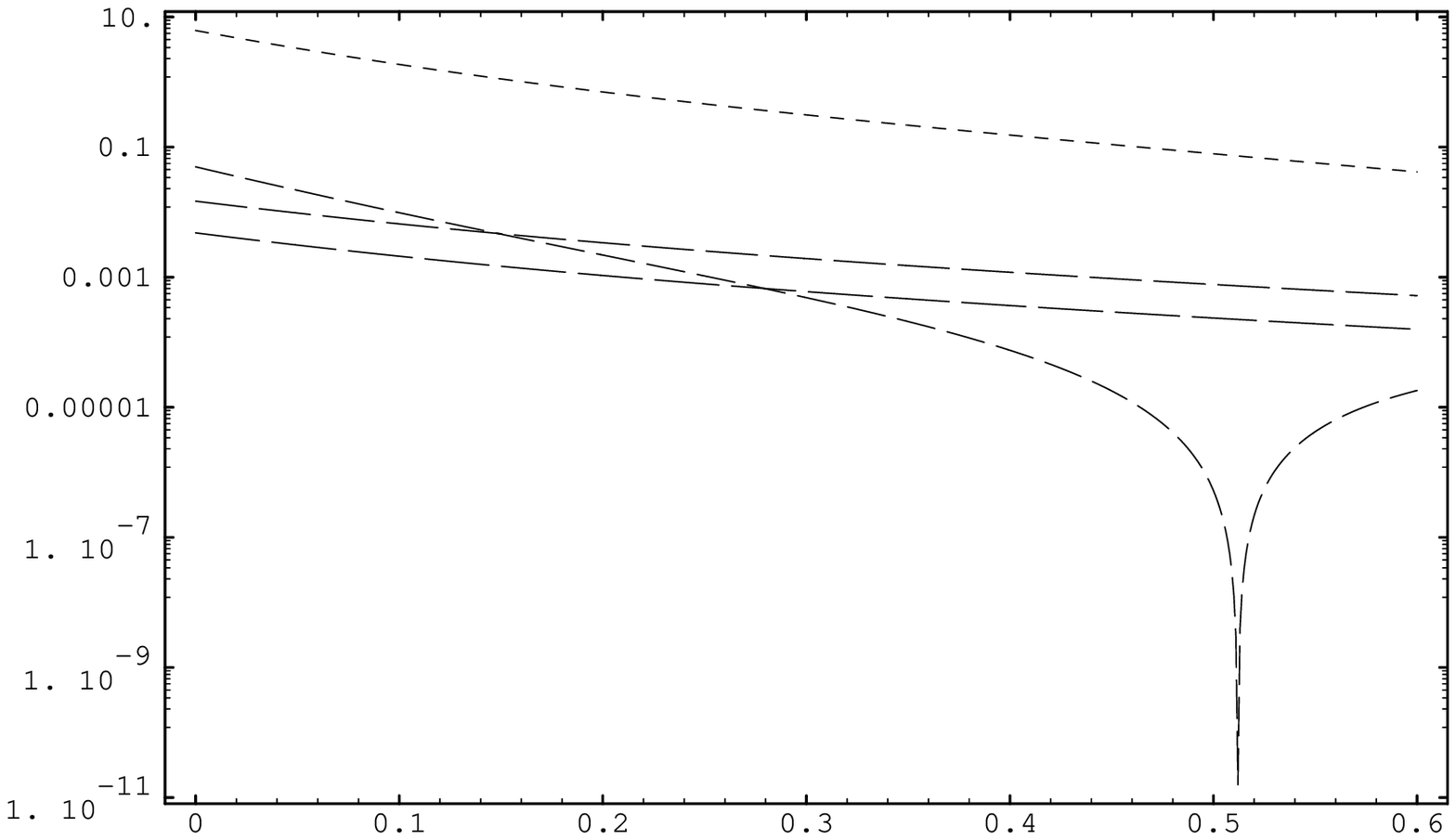}} }
  \put(67.25,45.5){\makebox(0,0){2S-Longitudinal}}
  \put(42,46){\makebox(0,0){\scriptsize $Q^2\!=\!1\!/\!4\GeV^2$}}
  \put(21,41){\makebox(0,0){\scriptsize $Q^2\!=\!2\GeV^2$}}
  \put(42,38){\makebox(0,0){\scriptsize $Q^2\!=\!10\GeV^2$}}
  \put(21.75,32.75){\makebox(0,0){\scriptsize $Q^2\!=\!20\GeV^2$}}
\end{picture}
\end{minipage}
\hspace*{-2mm}\begin{minipage}{81mm}
\setlength{\unitlength}{1mm}
\begin{picture}(80,49.5)
\put(-0.5,0){ \makebox(80,49.5){\figeps{80}{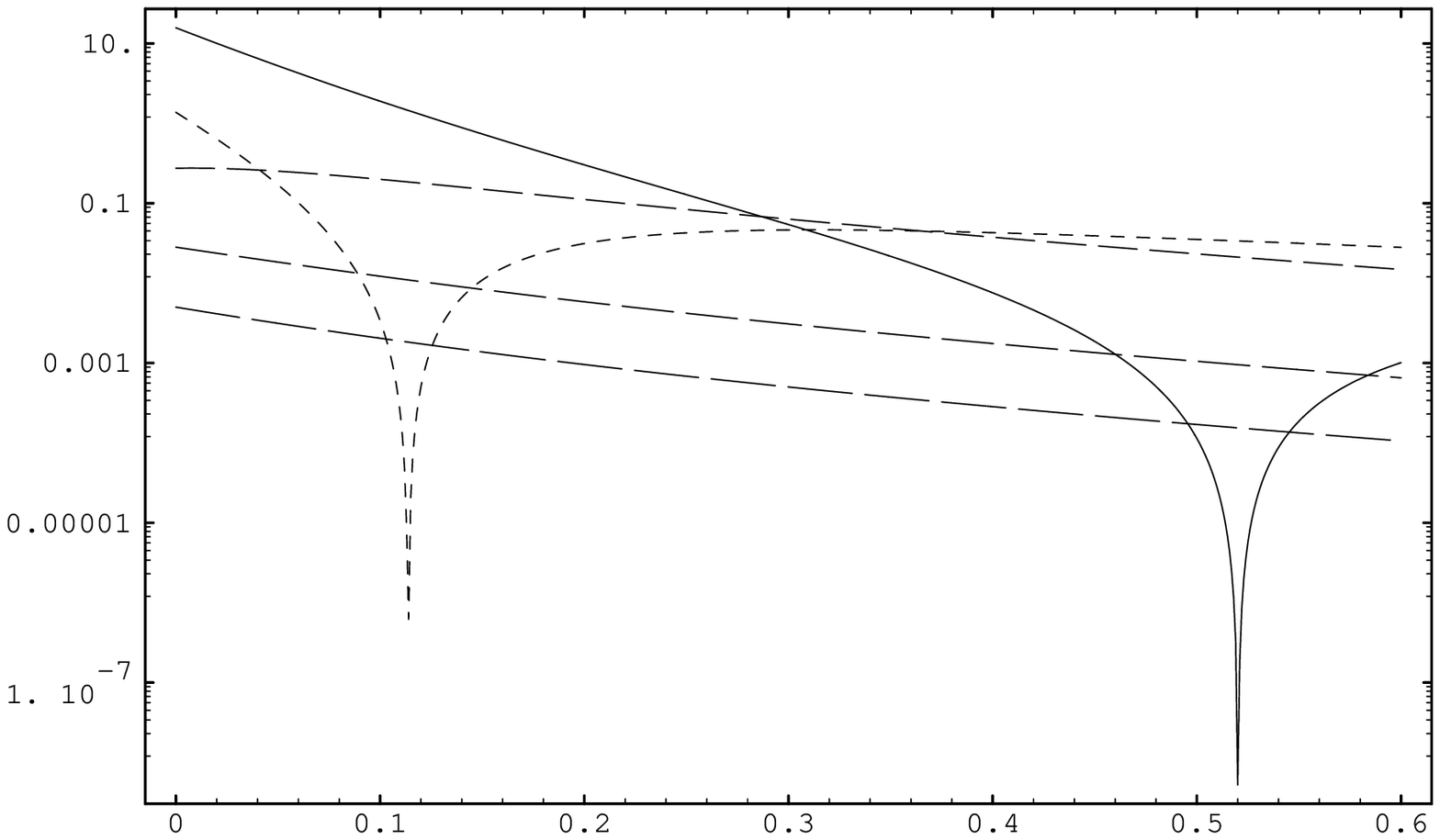}} }
  \put(74.25,-0.25){\makebox(0,0){$-t[\!\!\GeV^2]$}}
  \put(69,46.25){\makebox(0,0){2S-Transverse}}
  \put(64.75,7.5){\makebox(0,0){\scriptsize $Q^2\!=\!0$}}
  \put(20.5,11.5){\makebox(0,0){\scriptsize $Q^2\!=\!1\!/\!4\GeV^2$}}
  \put(72.5,32.75){\makebox(0,0){\scriptsize $Q^2\!=\!2\GeV^2$}}
  \put(42,34){\makebox(0,0){\scriptsize $Q^2\!=\!10\GeV^2$}}
  \put(42,25.5){\makebox(0,0){\scriptsize $Q^2\!=\!20\GeV^2$}}
\end{picture}
\end{minipage}
\vspace*{-5.5ex}
\caption[]{\footnotesize Differential cross sections as a function of $-t$ for the $\rh$-meson and the $2S$-state (upper and lower plots) for both longitudinal and transverse polarization (left and right). The curves with increasing dash sizes refer to \mbox{$Q^2\!=\!0,\; 1\!/\!4,\; 2,\; 10,\; 20\GeV$}. For the \mbox{$2S$-state} the node in the wave function has a strong influence on the $t$-dependence. } \label{Fig:dsdt}
\end{minipage}
\end{figure}

In Fig.~\ref{Fig:dsdt} we display our predictions for differential cross sections as a function of $-t$, the invariant momentum transfer squared. The curves follow roughly exponential behaviour with a slight upward curvature at larger values of $-t$. Due to the nodes in the wave functions for the $2S$-states in addition dips occur due to cancellations.

%
%
\begin{figure}
$$
\figepsSC{65}{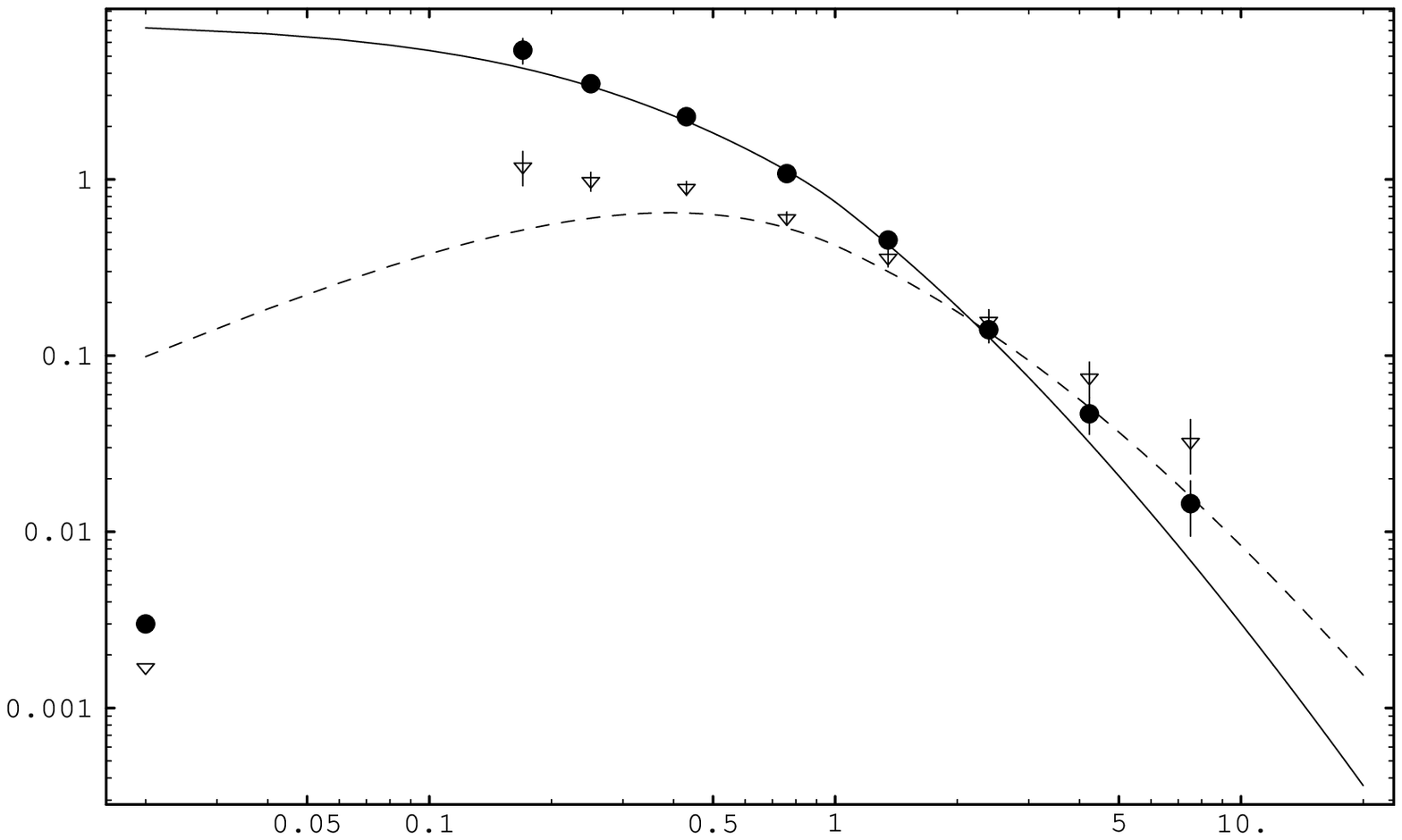}
\begin{picture}(0,0)
\put(-6,0){\makebox(0,0){$Q^2[\!\!\GeV^2]$}}
\put(-127,94){\makebox(0,0){$\si(Q^2)[\rm \mu b]$}}
\put(-109,80){\makebox(0,0){\scriptsize $\rh$-Transverse}}
\put(-106,50){\makebox(0,0){\scriptsize $\rh$-Longitudinal}}
\put(-96,27){\makebox(0,0){\scriptsize 85\% E665-Transverse}}
\put(-93,22){\makebox(0,0){\scriptsize 85\% E665-Longitudinal}}
\end{picture}
$$
\vspace*{-3ex}
$$
\figepsSC{65}{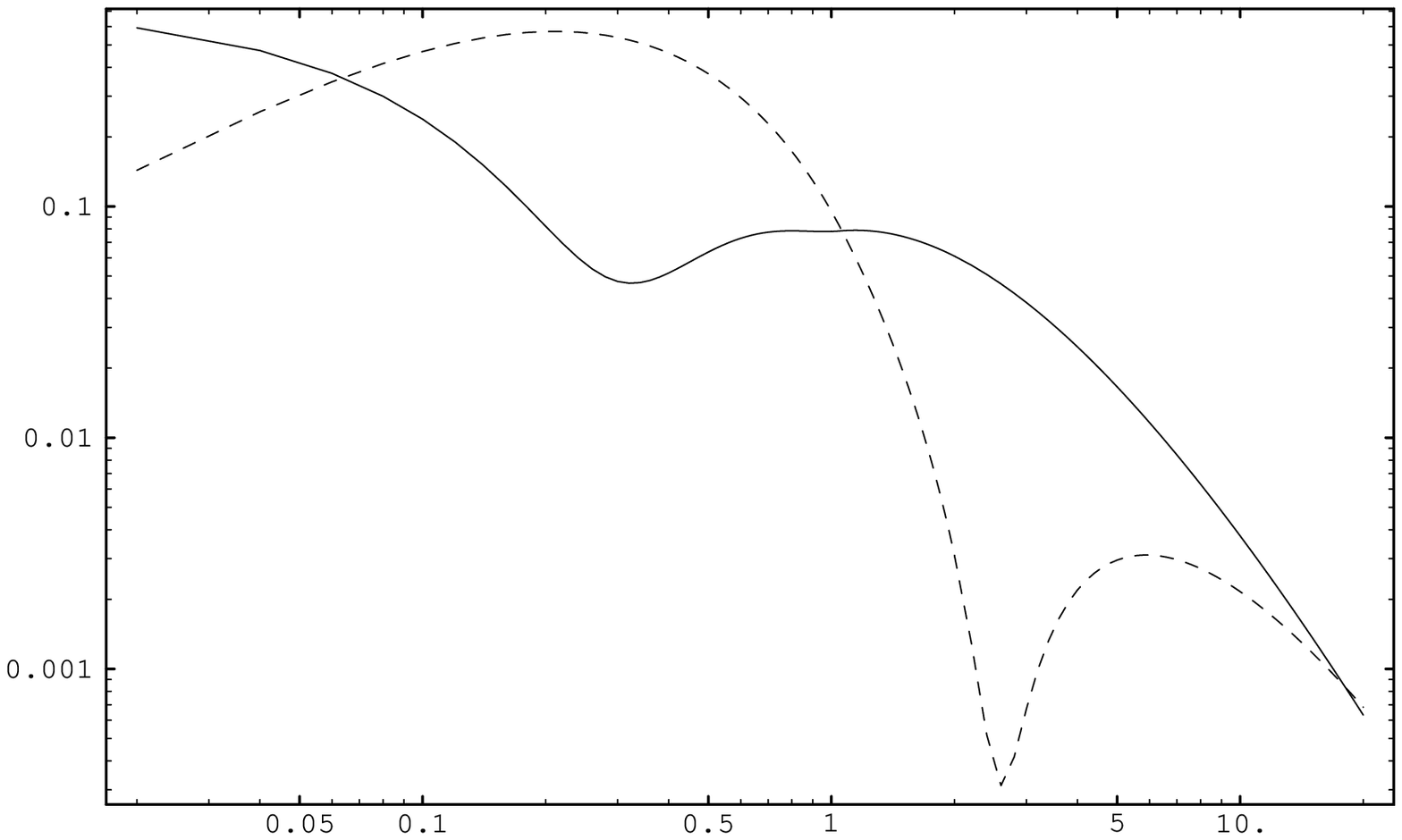}
\begin{picture}(0,0)
\put(-6,0){\makebox(0,0){$Q^2[\!\!\GeV^2]$}}
\put(-127,94){\makebox(0,0){$\si(Q^2)[\rm \mu b]$}}
\put(-40,80){\makebox(0,0){\scriptsize 2S-Longitudinal}}
\put(-21,65){\makebox(0,0){\scriptsize 2S-Transverse}}
\end{picture}
$$
\vspace*{-9.5ex}
\caption[]{\footnotesize Integrated cross sections of the $\rh$-meson and the $2S$-state as a function of $Q^2$. E665~\cite{E665_1} provides data for the $\rh$; we roughly estimate the pomeron contribution as 85\% of the measured cross section, cf. Ref.~\cite{DoLa}. } \label{Fig:sigma}
\end{figure}

Our results for integrated elastic cross sections as functions of $Q^2$ are given in Fig.~\ref{Fig:sigma}. For the $\rh$-meson our prediction is about $20-30\%$ below the E665-data~\cite{E665_1}. However, we agree with the NMC-experiment~\cite{NMC} which measures some definite superposition of longitudinal and transverse polarization, see Table~3 in Ref.~\cite{KDP}. For the $2S$-state we predict a marked structure due to the nodes of the wave function whose explicit shape, however, strongly depends on the parametrization of the wave functions.

\begin{figure}[!t]
$$
\figepsSC{65}{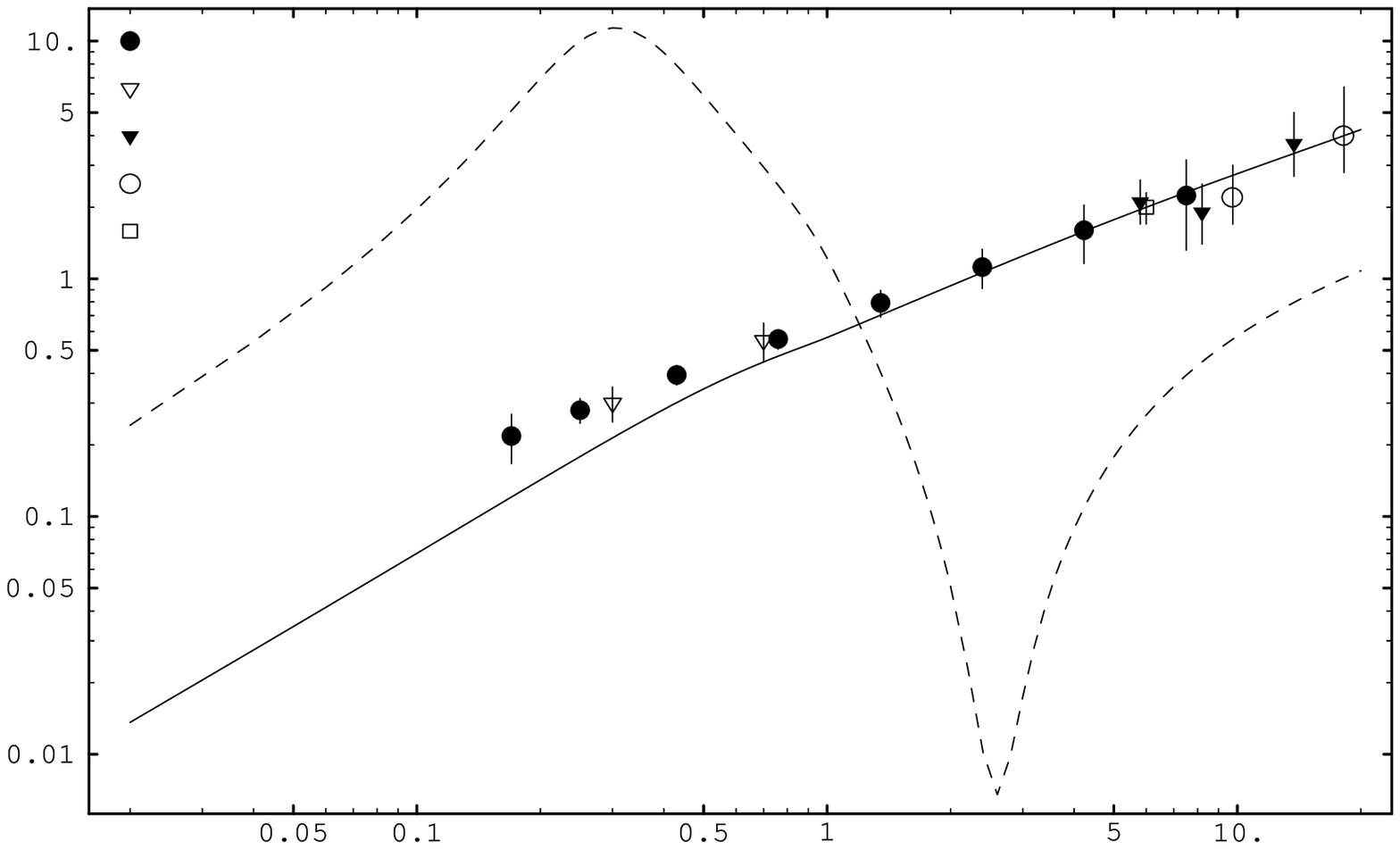}
\begin{picture}(0,0)
\put(-4,0){\makebox(0,0){$Q^2[\!\!\GeV^2]$}}
\put(-104,95){\makebox(0,0){$R_{LT}(Q^2)\!=\!\si^L(Q^2)\!/\!\si^T(Q^2)$}}
\put(-119,17){\makebox(0,0){$\rh$}}
\put(-117,47){\makebox(0,0){$2S$}}
\put(-118,85.75){\makebox(0,0){\scriptsize E665}}
\put(-95.5,81){\makebox(0,0){\scriptsize ZEUS BPC 1995 prel.}}
\put(-103,76.25){\makebox(0,0){\scriptsize ZEUS 1994 prel.}}
\put(-107,71.5){\makebox(0,0){\scriptsize H1 1994 prel.}}
\put(-117.5,66.75){\makebox(0,0){\scriptsize NMC}}
\end{picture}
$$
\vspace*{-10.5ex}
\caption[]{\footnotesize  Ratio of longitudinal to transverse integrated cross sections as function of $Q^2$ both for the $\rh$-meson and the $2S$-state. There is only data for $\rh$-production~\cite{E665_1}. } \label{Fig:R_LT}
\end{figure}

In Fig.~\ref{Fig:R_LT} we display the ratio $R_{LT}(Q^2)$ of longitudinal to transverse coss sections and find good agreement with experimental data for the $\rh$-state. For the $2S$-state we again predict a marked structure which is very sensitive to the node positions in the wave functions.
%
%

\begin{figure}[!b]
$$
\figepsSC{65}{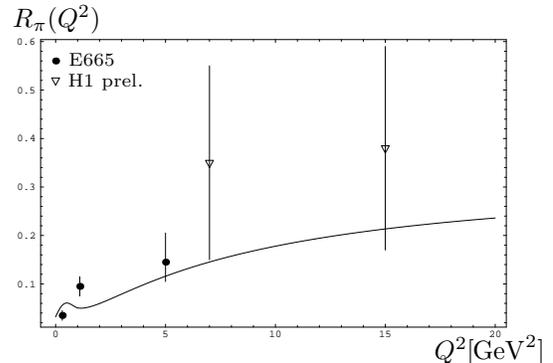}
\begin{picture}(0,0)
\put(-5,0){\makebox(0,0){$Q^2[\!\!\GeV^2]$}}
\put(-129,95){\makebox(0,0){$R_\pi(Q^2)$}}
\put(-119,83.5){\makebox(0,0){\scriptsize E665}}
\put(-115,77){\makebox(0,0){\scriptsize H1 prel.}}
\end{picture}
$$
\vspace*{-10.5ex}
\caption[]{\footnotesize Ratio of $2\pi^+2\pi^-$-production via $\rh'$ and $\rh''$ to $\pi^+\pi^-$-production via $\rh$ as function of $Q^2$. } \label{Fig:R_pi}
\end{figure}

Finally, in Fig.~\ref{Fig:R_pi} we confront with experimen\-tal data our calculation of the ratio $R_\pi(Q^2)$ of $2\pi^+2\pi^-$-production via $\rh'$ and $\rh''$ to $\pi^+\pi^-$-pro\-duction via~$\rh$ (for explicit definition see Ref.~\cite{KDP}).

%
%
\vspace*{-1.5mm}
%

%
%
\newpage
\renewcommand{\baselinestretch}{0.94}\normalsize
\section*{ACKNOWLEDGEMENTS}
The author wishes to gratefully acknowledge for collaboration in the underlying work H.G.~Dosch and H.J.~Pirner.

\end{document}